# Can life history predict the effect of demographic stochasticity on extinction risk?

Tobias Jeppsson[1]* & Pär Forslund[1]

[1] Department of Ecology, The Swedish University of Agricultural Sciences SLU, Box 7044, SE-75007 Uppsala, Sweden

* Corresponding author, email: Tobias.Jeppsson@slu.se, Tobias.Jeppsson@gmail.com

## Abstract

Demographic stochasticity is important in determining extinction risks of small populations, but it is largely unknown how its effect depends on the life histories of species. We modeled effects of demographic stochasticity on extinction risk in a broad range of generalized life histories, using matrix models and branching processes. Extinction risks of life histories varied greatly in their sensitivity to demographic stochasticity. Comparing life histories, extinction risk generally increased with increasing fecundity and decreased with higher ages of maturation. Effects of adult survival depended on age of maturation. At lower ages of maturation, extinction risk peaked at intermediate levels of adult survival, but it increased along with adult survival at higher ages of maturation. These differences were largely explained by differences in sensitivities of population growth to perturbations of life history traits. Juvenile survival rate contributed most to total demographic variance in the majority of life histories. Our general results confirmed earlier findings, suggesting that empirical patterns can be explained by a relatively simple model. Thus, basic life history information can be used to assign life history-specific sensitivity to demographic stochasticity. This is of great value when assessing the vulnerability of small populations.

# INTRODUCTION

Demographic stochasticity is an important intrinsic factor in determining the extinction risk of small populations (Soulé and Wilcox 1980, Shaffer 1981, Lande 1993, 1998). This stochastic process occurs because of the randomness inherent in the birth and death processes in a finite sample of individuals, and this will cause the realized population growth to deviate from the expected mean population growth rate (Lande 2002). The effect of demographic stochasticity has been studied theoretically (Richter-Dyn and Goel 1972, Keiding 1975, Leigh 1981, Mode and Pickens 1986, Gabriel and Burger 1992, Gilpin 1992, Lande 1993, Kokko and Ebenhard 1996, Lande et al. 2003, Engen et al. 2005) and empirically (e.g., Fujiwara and Caswell 2001, Sæther et al. 2004, Drake 2005, Melbourne and Hastings 2008) and is extensively used in population modeling (Beissinger and Westphal 1998, Caswell 2001, Beissinger and McCullough 2002, Morris and Doak 2004). Nevertheless, general relationships between characteristics of life history strategies and population dynamics caused by the effects of demographic stochasticity have rarely been investigated. If such relationships are significant, that is, if there are large differences between life histories in their sensitivity to demographic stochasticity, this is important information when assessing threat status and potential management actions for threatened species. Recognizing the life history of such species could thereby increase the accuracy in assessments of their extinction risk. The alternative, explicit species-specific population models, can be constructed for only a fraction of all threatened species due to lack of information. Therefore, management decisions must commonly be based on basic and often incomplete information of the biology of species, where rules of thumb that connect the biology of the species to extinction risk can be useful (Silvertown et al. 1996, Sæther and Bakke 2000).

A general model of stochastic population growth rate (log $\lambda_s$), accounting also for effects of environmental stochasticity on population growth rate, is

$$\log \lambda_s = r - \frac{1}{2}\sigma_e^2 - \frac{1}{2N}\sigma_d^2, \tag{1}$$

where $r$ is the mean population growth rate, $\sigma_e^2$ and $\sigma_d^2$ are environmental and demographic variances of population growth, respectively, and N is the population size (Lande 1998). Thus, the stochastic population growth rate is generally reduced by environmental and demographic stochasticity, and the effect of demographic stochasticity is reinforced as the population becomes smaller. Early models suggested that demographic stochasticity influences only very small populations, and has negligible effects when population sizes are larger than 100 individuals (Richter-Dyn and Goel 1972, Shaffer 1987, Lande 1993). It has since been acknowledged that the different demographic structures of life histories interact with demographic stochasticity. Ranges of population sizes over which demographic stochasticity is important therefore varies among species (Kokko and Ebenhard 1996, Fujiwara 2007), and may have significant effects even at population sizes of several thousands of individuals (Lande et al. 2003). Effects of demographic stochasticity may also depend on individual heterogeneity (Kendall and Fox 2002, 2003, Melbourne and Hastings 2008, Vindenes et al. 2008) and breeding system (Legendre et al. 1999, Gabriel and Ferrière 2004, Lee et al. 2011) but these factors are not considered here.

Interaction effects between demographic stochasticity and life history on extinction risk have only been studied to a limited extent. In theoretical models, Gilpin (1992) and Kokko and Ebenhard (1996) examined how some aspects of fecundity and delayed reproduction affect extinction risks. On the basis of empirical data from island birds, Pimm (1991) predicted that, at very small population sizes, species with short-lived and small individuals should be more adversely affected by demographic stochasticity than long-lived and large species. We only know of one empirical multispecies study of demographic stochasticity: in a study of birds, Sæther et al. (2004) found that 'fast' life-histories (i.e., low adult survival, large clutch size, and early maturation) had larger demographic variances than had 'slow' life histories. They reported a positive relationship between demographic variance and clutch size and negative relationships to age of maturation and generation time. Furthermore, their data suggested a curvilinear relationship between demographic variance and adult survival, indicating highest demographic variance at moderate values of survival (Sæther et al. 2004). However, the generality of these results has not been evaluated and a comprehensive theoretical account of how life-history traits relate to demographic stochasticity and extinction risk is still lacking.

In this article we model how demographic stochasticity affects extinction risk in different life histories. Life histories are explored along wide ranges of four life-history traits (adult and juvenile survival, fecundity, and age of maturation) included in partial life-cycle models (Caswell 2001, Oli and Zinner 2001, Oli 2003b). Using matrix modeling, we apply branching processes to determine extinction risk (Caswell 2001, Fujiwara 2007). We thereby offer a framework within which empirical results, such as those of Sæther et al. (2004), can be fitted. Extinction is caused by demographic stochasticity alone in our study. To understand the population dynamic processes that result in the life-history-related patterns of extinction risk, we compute total demographic variance of population growth and the specific contributions to demographic variance from the different life-history traits (Engen et al. 2005).

## METHODS

To model life histories we used partial life-cycle analysis (Caswell 2001, Oli and Zinner 2001), where matrix population models are constructed from basic demographic data. We assume a constant environment and no density-dependence. To minimize the number of parameters in the model we used a simplified version in which age at last reproduction is omitted. The model only consists of four parameters (Oli 2003a). The life history traits included in our model are juvenile survival ($P_j$ - defined as survival of any prereproductive class), adult survival ($P_a$), fecundity ($m$) and age of maturation ($a$). All stages in the model where individuals have not reached maturity are termed juvenile stages. We modeled a post-breeding census, so the fertility terms ($F_x$) in the matrix model (eq. [1]) are calculated as $P_x \times m$, where $x$ denotes stage class $x$. Note that throughout the paper we will refer to $m$, that is, number of offspring, as fecundity, and $F_x$ as fertility. The choice of a post-breeding model is arbitrary, but is useful since demographic information for many vertebrates is collected this way. Our modeling approach allows for a large range of life histories to be explored by varying only a few life history traits. The model is a single sex model of the form

$$\mathbf{A} = \begin{bmatrix} 0 & 0 & \cdots & F_j & F_a \\ P_j & 0 & \cdots & 0 & 0 \\ 0 & P_j & \cdots & 0 & 0 \\ \vdots & \vdots & \ddots & \vdots & \vdots \\ 0 & 0 & \cdots & P_j & P_a \end{bmatrix}, \qquad N_{t+1} = \mathbf{A} N_t. \tag{2}$$

Matrix **A** is the matrix describing the expectation of survival and reproduction for a particular life history strategy, with a dominant eigenvalue representing the mean population growth rate ($\lambda = e^r$) and associated left and right eigenvectors representing reproductive value (**v**) and stable stage distribution (**w**), respectively. $F_j$ ($= P_j \times m$) is fertility of the last juvenile stage class and $F_a$ ($= P_a \times m$) is fertility of the adult stage class, as follows from standard procedures of the post-breeding census method in matrix population modeling (Caswell 2001). The dimension of the matrix (number of rows and columns) is determined by the age of maturation and is ($\alpha+1$, $\alpha+1$). Below, matrix entries (e.g., $P_j$, $F_a$) are sometimes referred to as $a_{ij}$ as the entry in the $i$th row in the $j$th column.

To study extinction risk of different life histories we explored a range of parameter values for the four parameters in the model, where each combination of life-history traits represents a hypothetical species. We chose to model ranges of semelparous to iteroparous life histories (range of $P_a$: 0.0–0.95, steps of 0.025), few to many offspring (range of $m$: 0.5–30; 0.5, 1, 1.5, 2, and then in steps of 1 up to 30), and early to late maturation (range of $\alpha$: 1–15, steps of 1). Our method of constructing life histories generates some life-history combinations that are rare or maybe non-existent in nature due to different kinds of constraints (e.g., physiological or environmental constraints). Nevertheless, we still include these life histories in our analysis because they are important for the generic understanding of the effect of demographic stochasticity on population dynamics.

The mean population growth rate is a major determinant of population extinction risk over a specific time interval (Caswell 2001, Sæther et al. 2005), but we wanted to analyze the specific effects of life history on demographic variance and extinction risk. We therefore standardized the population growth rate so that $\lambda = 1$ for all life histories (similar to Kokko and Ebenhard 1996, Jonsson and Ebenman 2001, Fujiwara 2007, Lee et al. 2011). Further, we assumed that there was no environmental variation. Relating to equation (1), this means that $r = 0$ and $\sigma_e^2 = 0$. We could thereby investigate the specific effects of demographic stochasticity, and make relative comparisons of extinction risks and demographic variances between life histories. To create life histories, we treated three of the parameters in the model as fixed and calculated the fourth (chosen to be $P_j$) to meet the criterion $\lambda = 1$. Solving the characteristic equation of A (Caswell 2001) for this parameter results in $P_j = [m/\lambda^\alpha + P_a m/(\lambda^{\alpha+1} - P_a \lambda^\alpha)]^{-1/\alpha}$. Since we assumed that $\lambda = 1$, the characteristic equation reduces to:

$$P_j = \left(m + \frac{P_a m}{1 - P_a}\right)^{-1/\alpha}. \quad (3)$$

Thus, $P_j$ will decrease with increasing $m$ and $P_a$, and will increase with increasing $\alpha$. Equation (3) can yield survival values larger than one, so we set the maximum juvenile survival to 0.99. With this restriction, $\lambda = 1$ cannot be achieved for some combinations of adult survival, fecundity and age of maturation. Life histories corresponding to those cases were removed from the analysis. In the analysis of each of the traits (adult survival, fecundity and age of maturation), we set up different combinations of the two other traits (see figs. 1–3). This yielded a total of 1012 different life histories that are presented in figures 1–3 and appendix B, available online. To estimate how the extinction risk of different life histories was affected by demographic stochasticity we used the mathematical framework of branching processes, more specifically multitype branching processes (Harris 1963, Caswell 2001). Starting with a population of $n$ individuals at time $t_0$, each individual can give rise to 'offspring' in each time step (in this terminology 'offspring' includes a surviving parent and its true offspring), based on probability distributions of reproduction, survival and transitions. These offspring can in turn produce offspring in the following time steps. Therefore, each individual in the starting population gives rise to a tree of descendants, that is, an individual trajectory. At any point in time ($t$) all descendants in an individual trajectory may go extinct, and the whole population goes extinct when all individual trajectories have gone extinct. Branching processes can be used to calculate the analytical extinction risk at time $t$ of individual trajectories from the starting population, which can be collated to calculate the extinction risk of the entire population. The probability of individual trajectory extinction is determined by probability generating functions, which can be derived from a matrix model (Caswell 2001, Fujiwara 2007), given assumptions regarding distributions that govern survival and reproduction. We have used a rather simple approach and treated survival probabilities ($P_x$) as binomially distributed and fertilities ($F_x$) as Poisson distributed (identical to Fujiwara 2007; see also Caswell 2001, and Morris and Doak 2002). Variances are $P_x(1-P_x)$ for survival rates, and equal to the mean $F_x$ for fertilities.

To compare life histories we calculated extinction risks ($Q_t$) over 100 years for a total initial population size of 100, starting at their stable stage distribution (**w**), as

$$Q_t = \prod_i (q_{i,t})^{\mathbf{n}_i}. \quad (4)$$

Term $Q_t$ is the risk of population extinction over t years, where $q_{i,t}$ is the trajectory extinction risk of stage $i$ at time $t$ as estimated from the probability generating function (derived from the branching process) and **n** is the initial population vector (Caswell 2001).

Demographic variances of population growth ($\sigma_d^2$) and trait-specific contributions to $\sigma_d^2$ were estimated according to Engen et al. (2005):

$$\sigma_d^2 = \sum_{i=1}^{k} w_i^{-1} \left(s_{Fi}^2 \sigma_{Fi}^2 + s_{Pi}^2 \sigma_{Pi}^2 + 2 s_{Fi} s_{Pi} \tau_i \right), \quad (5)$$

where $\mathbf{w}_i$ is the proportion of individuals in stage $i$ at stable stage distribution, $\sigma^2_{Fi}$ and $\sigma^2_{Pi}$ are variances of fertility ($F_i$) and survival ($P_i$) at stage $i$, respectively, and $k$ is the number of life stages (i.e., age of maturation + 1 in our case). Term $\tau_i$ is the covariance of reproduction and survival of stage $i$, assumed to be zero in this study. Finally, $s_{Fi}$ and $s_{Pi}$ are sensitivities of population growth rate to perturbations of the matrix entries fertility and survival at stage $i$, respectively (Caswell 2001). Sensitivity for the matrix entry $a_{ij}$ (see eq. [2]) was calculated as $\delta\lambda/\delta a_{ij} = \mathbf{v}_i\mathbf{w}_j$, where the reproductive value $\mathbf{v}$ was adjusted so that the scalar product of the left ($\mathbf{v}$) and right ($\mathbf{w}$) eigenvectors of $\lambda$ was 1, that is, $\langle\mathbf{w},\mathbf{v}\rangle = 1$ (for further details, see Caswell 2001). Sensitivity measures how much $\lambda$ will change if a matrix entry change by a fixed amount. Thus, sensitivities measure the degree to which variation in matrix entries due to demographic stochasticity will cause variation in population growth, that is, total demographic variance. Note, though, that the sensitivities we calculated are for matrix entries, which means that reproduction is represented by fertility ($F_j = P_j \times m$) and not fecundity ($m$). To simplify the presentation we will commonly refer to sensitivities as, for example, sensitivity of fertility, although we have the above definition in mind (but see Caswell 2001, p. 208).

Using equation (5) we calculated the contribution of demographic variance from each matrix entry to the total demographic variance. To calculate the total contributions from juvenile survival and fertility, we summed the stage specific contributions of these parameters from all stages. Ranges of values of life histories traits used for calculation of total demographic variances in figure 4 are described in the figure text. For estimating contributions to total demographic variance, we created life histories from all combinations of the ranges of life history trait values listed above. This resulted in a total of 18,367 life histories fulfilling the criterion $\lambda = 1$. This set of life histories is used in the analysis of contributions to total demographic variance (see Results), although only a selection of the life histories are shown in figures 5, A2 and A3.

## RESULTS

### LIFE HISTORY SPECIFIC EFFECTS OF DEMOGRAPHIC STOCHASTICITY

To gauge how demographic stochasticity affects a spectrum of life histories, we calculated extinction risks over ranges of the life-history traits fecundity, adult survival, and age of maturation (full data found in appendix B). It should be kept in mind that in all of these calculations, holding $\lambda = 1$ causes juvenile survival to decrease with increasing fecundity or adult survival, and to increase with increasing age of maturation. The extinction risk due to demographic stochasticity showed distinct differences among life histories. In the majority of cases, extinction risk increased with increased fecundity (fig. 1). An exception was found for life histories with early maturation and low adult survival, where extinction risk actually decreased with increasing fecundity (fig. 1A).

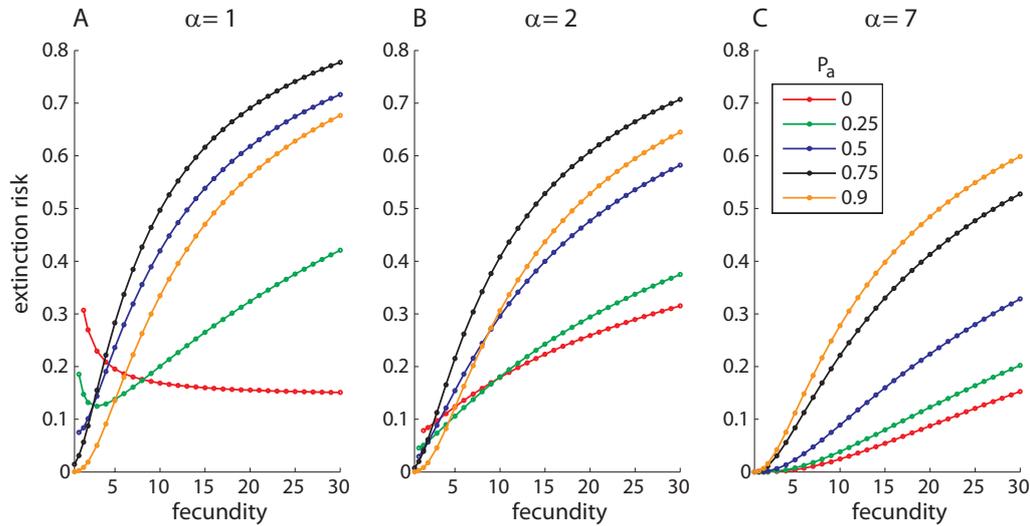

**Figure 1**: Extinction risk after 100 years due to demographic stochasticity for life histories with different fecundities (shown on X-axis), different adult survival rates (shown as lines, see legend for values) and different ages of maturation (α; A–C). Each life history (see app. B, available online) is indicated by a dot. Initial population size $N_0 = 100$, mean population growth rate $\lambda = 1$ (see "Methods" for details).

The effects of adult survival were more complex, although extinction risk generally peaked at intermediate levels of adult survival (fig. 2). Adult survival interacted, however, with age of maturation and fecundity. For life histories with low age of maturation and moderate to high fecundity, extinction risk peaked at intermediate adult survival rates, whereas extinction risk was a decreasing function of adult survival at low fecundity. Extinction risk was generally higher at higher fecundities (figs. 2A, 2B). In life histories with high ages of maturation, extinction risk increased with increased adult survival up to $P_a = 0.9$ (fig. 2C).

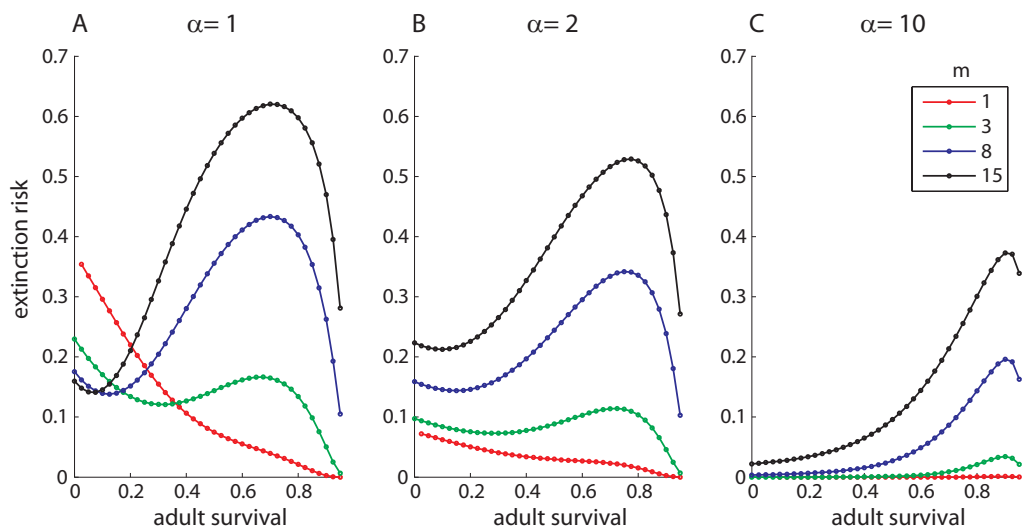

**Figure 2:** Extinction risk after 100 years due to demographic stochasticity for populations with different adult survival rates (shown on X-axis), different fecundities (shown as lines, see legend for values) and different ages of maturation (α; A–C). Each life history (see app. B, available online) is indicated by a dot. Initial population size $N_0 = 100$, mean population growth rate $\lambda = 1$ (see "Methods" for details).

Age of maturation had a strong effect on extinction risk in life histories with low to moderate adult survival. Here, extinction risk was a steep decreasing function of age of maturation in life histories with low adult survival and moderate adult survival (fig. 3A,B). The effect of age of maturation on extinction risk for semelparous life histories ($P_a = 0$) was almost identical to the case $P_a = 0.1$ (semelparous not shown in fig. 3, but cf. fig. 2). At higher adult survival, extinction risk was only slightly negatively related to age of maturation at high fecundities, and almost unrelated to age of maturation at low fecundities (fig. 3C). An exception to the negative relationship between age of maturation and extinction risk was found for life histories with low adult survival and low age of maturation, where extinction risk increased going from an age of maturation of 1-2 years (fig. 3A).

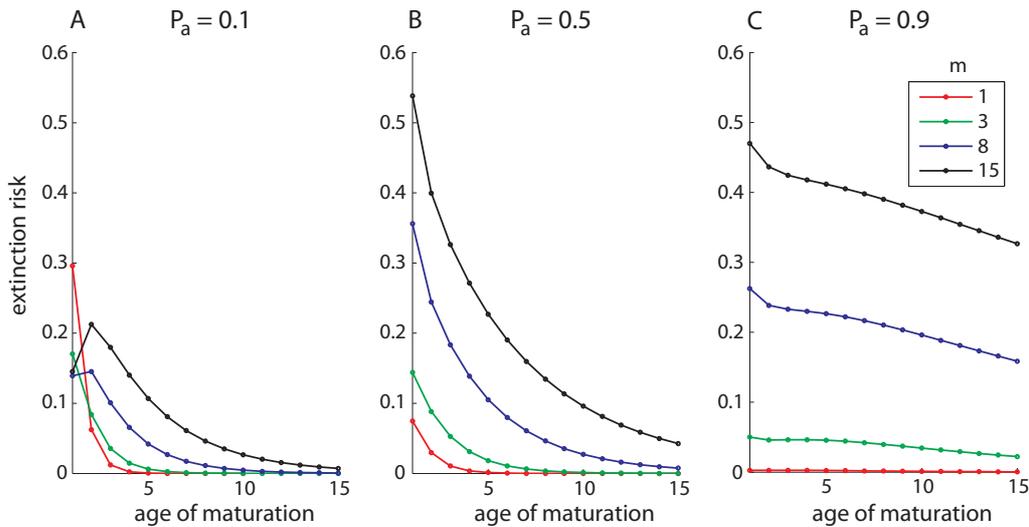

**Figure 3:** Extinction risk after 100 years due to demographic stochasticity for life histories with different age of maturation (shown on X-axis), different fecundities (shown as lines, see legend for values) and different adult survival rates ($P_a$; A–C). Each life history (see app. B, available online) is indicated by a dot. Initial population size $N_0 = 100$, mean population growth rate $\lambda = 1$ (see "Methods" for details).

We examined the robustness of our results to the assumption that $\lambda = 1$ by changing it in the range 0.95–1.05, and by also using initial population sizes of 1000 (when $\lambda < 1$) and 20 (when $\lambda > 1$). Here, we used the general form of equation (3) to create life histories since $\lambda \neq 1$. This simulated cases of threatened populations that decrease towards critically low population sizes or recover from such situations. Although the absolute levels of extinction risk of course increased or decreased, patterns of extinction risk between life histories remained (not shown). This confirmed that our results were qualitatively robust.

Extinction risk depends on population size and the time period used. To relax those constraints and widen the generality of our results, we visualized the relationships between the life-history traits and total demographic variance (eq. [5]), which are independent of these factors. This showed, as expected, that the results described above (figs. 1–3) were similar to those found for demographic variance (fig. 4), and that there was a strong and positive sigmoidal relationship between extinction risk and total demographic variance (fig. A1).

Total demographic variance generally decreased with increasing juvenile survival (fig. 4), which means that also extinction risk had the same relationship with juvenile survival (cf. fig. A1). At medium to high juvenile survival rates, demographic variance was generally low, almost irrespective of age of maturation (figs. 4C–4F). At low to very low juvenile survival rates, on the other hand, demographic variance peaked at high fecundity and survival rates around 0.75 at ages at maturation of one and two years (figs. 4A, 4B).

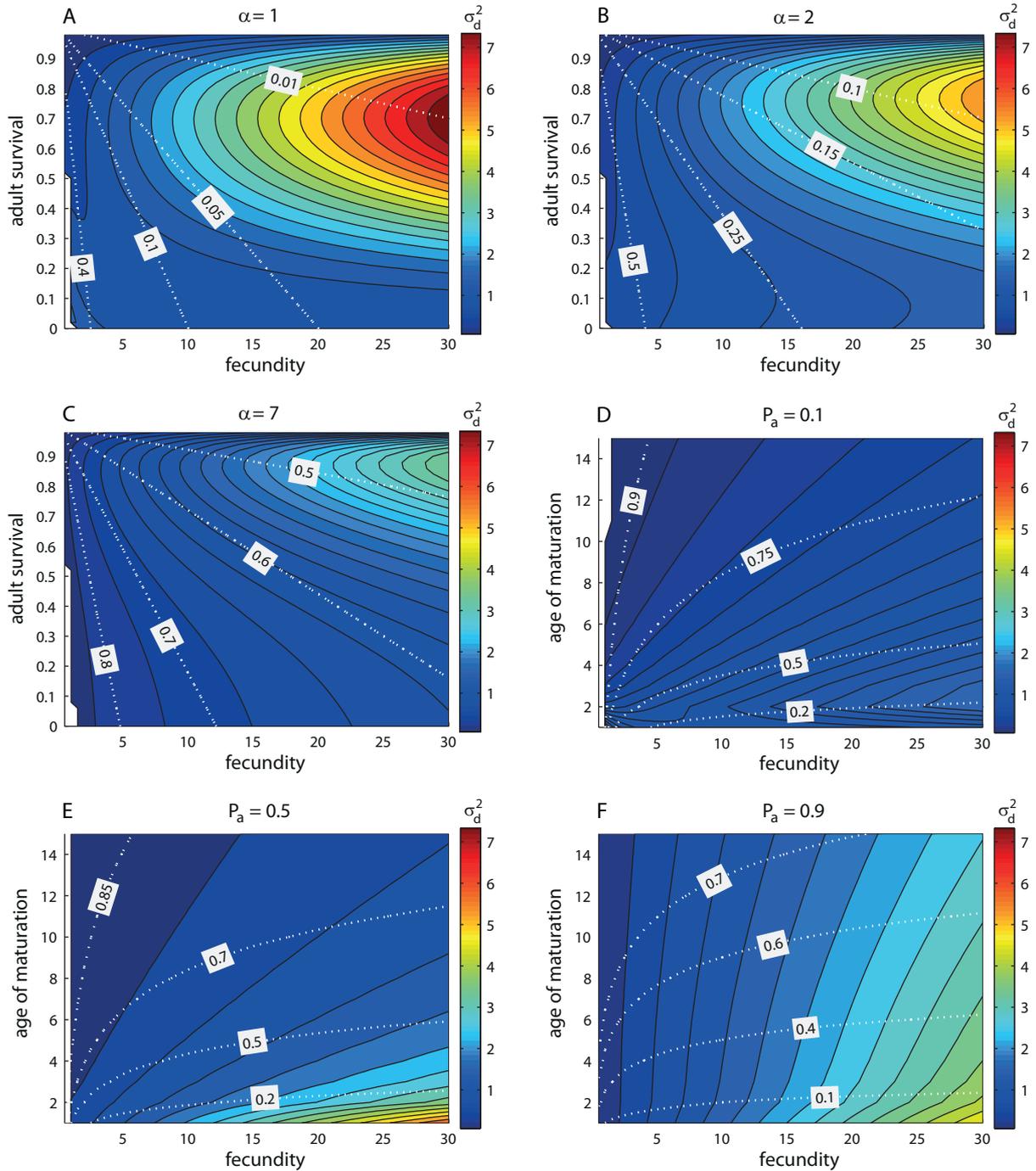

**Figure 4:** Contour graphs showing the demographic variance over ranges of life histories, as calculated from equation (5). The legend to the right shows the color scale of values of demographic variance. Graphs A–C show life histories with adult survival (range = 0–0.98; steps of 0.02) and fecundity (range

= 0.5–30; steps of 0.5) at age of maturation of 1 (A), 2 (B) and 7 years (C). Graphs D–F show life histories with age of maturation (range = 1–15; steps of 1) and fecundity (range = 0.5–30; steps of 0.5) at adult survival equal to 0.1 (D), 0.5 (E) and 0.9 (F). Juvenile survival rates, estimated from equation (3), are superimposed on the graphs as isoclines where labels show the juvenile survival rate. White areas of the graphs represent life histories where the expected population growth rate could not be standardized to one, given our model assumptions.

### DEMOGRAPHIC VARIANCE COMPONENTS

The general findings that extinction risk increased with increased fecundity, peaked at intermediate survival and decreased with increasing age of maturation (figs. 1–3), were generally explained by contributions of demographic variance from juvenile survival to the total demographic variance (fig. 5). Although figure 5 only shows a selection of life histories, the same result was found for the majority of the other life histories as well. To investigate which stage specific survival and fertility term that contributed most to total demographic variance in the range of analyzed life histories, we compared contributions among life histories (N = 18,367). This showed that the contribution from juvenile survival was largest in c. 94% of the life histories. Demographic variance contributions from fertility ranked highest in c. 6% of the cases, and adult survival contributions in c. 0.1%. In nine cases juvenile and adult survival rate both ranked highest, and in one case juvenile survival rate and fertility both ranked highest. Despite this, it was clear that juvenile survival rate contributed most to the total demographic variance in the majority of cases. However, the contribution from fertility ranked highest for life histories that had low adult survival, low age of maturation and low fecundity, that is, life histories where we found exceptions to the general patterns (Figs. 1–3, 5). Demographic variance contributions from adult survival rate ranked highest for some long-lived life histories with low fecundity (not shown).

Demographic variance contributions are functions of sensitivities and variances of stage specific survival and fertility (eq. [5]). Decomposing the contributions further into these components revealed that sensitivities had a major effect on the demographic variance contributions, which can be explained by the quadratic effect of sensitivity in equation (5). Variance in stage specific survival and fertility (assumed to be binomially and Poisson distributed, respectively), on the other hand, commonly did not explain the patterns of total demographic variance. An exception to this was for the drop in extinction risks in life histories with very high adult survival rates (fig. 2). Here, both sensitivities and variances of juvenile and adult survival rates had significant impacts on shaping the pattern (figs. 5D–5F, A2D–A2F, A3D–A3F). Thus, although the variance of fertility increased with increasing mean as a consequence of using the Poisson distribution (figs. A3A–A3C), the demographic variance contribution from fertility decreased with increasing fecundity (figs. 5A–5C). The reason for this was that increased fecundity causes a larger proportion of the population to be juvenile individuals, which yields lower sensitivities to fertility in high than in low-fecundity life histories (because fertility sensitivities = $\mathbf{v}_1 \times \mathbf{w}_i$, where $\mathbf{v}_1$ is the reproductive value of the first life stage; see eq. [5] and Caswell 2001).

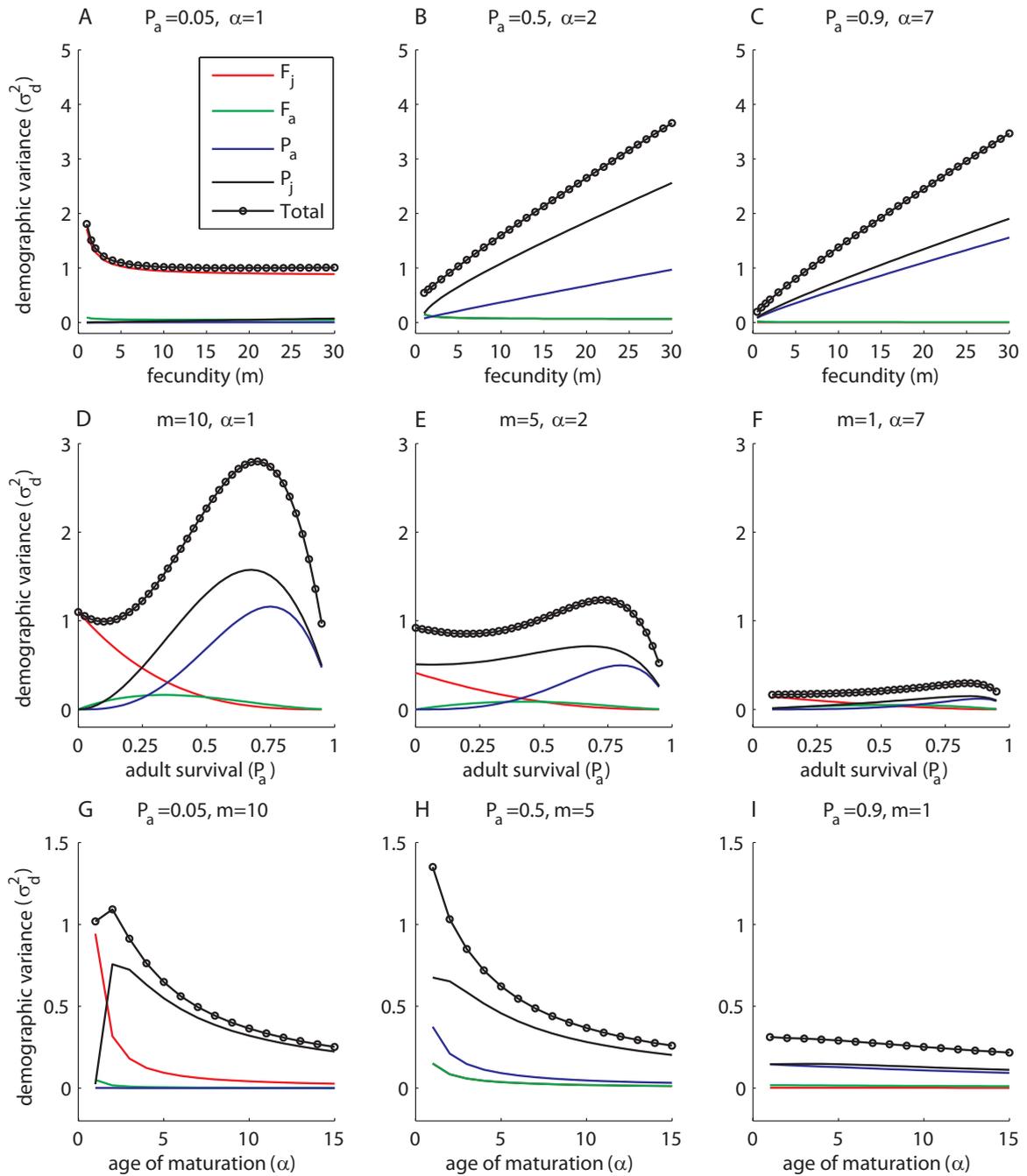

**Figure 5:** Total demographic variance of life histories and contributions to total demographic variance from different matrix entries in relation to ranges of the life history traits. Total demographic variance and the contributions are estimated from equation (5) (see "Methods"). The figures (A–I) show a selection of the total array of life histories investigated. Titles above each graph states which life history traits that are fixed. A legend common to all graphs is shown in the first graph.

Peaks in extinction risk and total demographic variance at intermediate adult survival rates were caused mainly by sensitivities for juvenile and adult survival rates, although declining variances for adult survival rate caused the contribution from this trait to total demographic variance to decline steeply at high adult survival rates (figs. 5D–5F, A2D–A2F, A3D–A3F). The shift in the peak of total demographic variance (and hence extinction risk) in relation to adult survival rate with increasing age of maturation was due to different levels of juvenile survival rates at different ages of maturation (fig. 4).

This affected sensitivities and variances of juvenile survival rate (figs. A2D–A2F, A3D–A3F). Variance of juvenile survival rate decreased with increasing adult survival rate at low ages of maturation, but increased at high ages of maturation (figs. A3D–A3F) owing to whether juvenile survival rate approached or moved away from $P_j = 0.5$, where maximum variance is attained. These differences in variance combined with sensitivities explain the shift in the peak of total demographic variance and extinction risk (fig. 2).

The general decline in total demographic variance with increasing age of maturation was caused by decreasing contributions from all stage specific survival and fertility rates (figs. 5G–5I). Decreasing contributions were, in turn, caused by decreasing sensitivities (figs. A2G–A2I). Variances could not explain the general decline (figs. A3G–A3I). The decrease of sensitivities occurred because the components of sensitivity, that is, reproductive value and stable stage distribution, were distributed more evenly among life stages as age of maturation increased (not shown).

## DISCUSSION

We modeled extinction risks caused by demographic stochasticity for a large spectrum of life histories. The life history dimensions covered are early to delayed reproduction, few to many offspring, and semelparous to iteroparous reproduction. The general modeling approach is broadly applicable to a large range of organisms (Oli 2003a), such as birds, mammals, fish (although somatic growth after maturation is not considered), arthropods with ≥ 1 year life cycles, and some plants with ≥ 1 year life cycles. Overall, extinction risks increased with fecundity and decreased with age of maturation. An exception to this general pattern was for semelparous life histories within a range of low ages of maturation, where extinction risk decreased with fecundity and increased with age of maturation. With respect to adult survival, extinction risk peaked at intermediate levels of adult survival for most life histories, whereas it was negatively or positively related to adult survival in others. Overall we found that the specific effects of life history traits on extinction risk depended on the whole life history of the organism; that is, life history traits interacted in determining extinction risk. Our results therefore confirm and expand the results of Kokko & Ebenhard (1996), showing that extinction risk may vary among life histories due to demographic stochasticity. Furthermore, our results partially support the general hypothesis that 'fast' life histories are more susceptible to effects of demographic stochasticity more strongly than 'slow' life histories (Gilpin 1992).

A positive relationship between extinction risk and fecundity due to demographic stochasticity has been established previously (Gilpin 1992, Kokko and Ebenhard 1996). Thus, among long-lived organisms, one would expect geese and crocodilians (which produce many offspring) to be more sensitive to demographic stochasticity than albatrosses or elephants (which produce few offspring). We further showed that the effect of fecundity interacts with adult survival and age of maturation in shaping extinction risks, so that the reverse relationship can be true in some situations, as in very short-lived life histories (fig. 1A). Here, a short-lived life history with low fecundity had higher extinction risk than life histories with higher fecundities. However, such a low fecundity life history may be difficult to find in reality although cavies have life histories with some similarities to this (Kraus et al. 2005). Rather, a result of more practical interest regarding short-lived life histories is that extinction risk seemed to be similar among those that had higher fecundities (fig. 1A, $P_a = 0$, $m > 5$), that is, fecundity was not important in determining extinction risk in these life histories.

Extinction risk decreased with increased age of maturation, although for the most long-lived species this relationship was weaker. Therefore, delayed reproduction (coupled to higher juvenile survival) generally seemed to lower the risk of extinction. This difference was due to overall lower contributions to total demographic variance from mainly juvenile survival but also from adult survival and fertility (figs. 5G–5I). Life histories with late maturation and high adult survival, fecundity and juvenile survival still had, however, high extinction risks. Thus, late maturation does not always buffer against extinction. Kokko & Ebenhard (1996), using a similar modeling approach, showed that the effect of delayed reproduction on demographic stochasticity depended on the level of fecundity. They found that a delayed life history had higher extinction risk as compared to a nondelayed one when both had high fecundity, whereas the opposite was true at low fecundity. Our study confirmed these results, but also showed that this relationship is valid only for life histories with low adult survival and low age of maturation (fig. 3A).

In general, the extinction risk peaked at intermediate adult survival rates, but this pattern was molded by age of maturation, fecundity and juvenile survival. Thus, extinction risk decreased with increasing adult survival in life histories with low fecundity and low age of maturation. Further, in life histories with late maturation, extinction risk increased with adult survival rate. This is because increasing age of maturation pushes the extinction risk peak to higher levels of adult survival due to changing variances and sensitivities of juvenile and adult survival (figs. 2, A2D–A2F, A3D–A3F), and for a sufficiently high age of maturation the maximum extinction risk is at an adult survival rate of 1. If these results are framed in terms of semelparous and iteroparous life histories, the interpretation is that semelparous life histories often have lower extinction risks than have iteroparous ones. One example of this is in salmonids, where Fujiwara (2007) estimated extinction risks due to demographic stochasticity to be lower in two semelparous species (chinook and coho salmon) than it was for an iteroparous species (steelhead salmon). Such differences between semelparous and iteroparous species seem to be due to higher contributions to demographic variance from adult and juvenile survival in the iteroparous life history (fig. 5D). The negative effects of iteroparity were, however, reduced by earlier maturation and lower fecundity.

The differences in extinction risk were caused by differences in total demographic variance of population growth between life histories. The total demographic variance is a result of demographic variance contributions from different stage specific survival and fertility rates. Our analysis showed that these contributions were primarily determined by the sensitivities of population growth to changes in stage specific survival and fertility. Variance in these rates, generated by demographic stochasticity, could not alone explain the relationships between extinction risks and life-history strategies. This implies two important things. First, our choice of probability distributions of life history traits in the model was not critical to the relationships we found between extinction risk and life history traits. Thus, variances of stage specific survival and fertility act mostly as scaling factors, whereas sensitivities create the patterns of the demographic variance contributions. For example, assuming probability distributions for fecundity other than the Poisson distribution (Kendall & Wittman 2010) may change the levels of the demographic variance contributions from fertility but likely not the general relationship with fecundity.

Second, sensitivities explain much of the stochastic dynamics of different life histories due to demographic stochasticity. This means that the relationships between extinction risk and life-history traits are caused by population structure. Further, sensitivities can be broken down into stable stage distributions and reproductive values (Caswell 2001). For example, in our modeling framework the proportion of individuals that are in the first stage class (i.e., juveniles) will be larger in high-fecundity life histories than in low-fecundity life histories, whereas the reproductive value of the first stage class will be relatively lower in high-fecundity than in the low-fecundity life histories (cf. Charlesworth 1994, Caswell 2001). The same effects occur when comparing life histories with low adult survival and high adult survival. However, the resulting effects of these differences between life histories on sensitivities are not intuitive, but depend on how large the differences are in stable stage distribution and reproductive value, respectively, and how they combine in sensitivities. Understanding the integrated effects of stable stage distribution and reproductive value on sensitivities can therefore generally explain differences in stochastic dynamics between life histories. A general theoretical analysis of this interaction, focusing on the differences between different kinds of life histories, is lacking, to our knowledge, but should greatly increase our understanding of extinction dynamics of different life histories.

In the vast majority of the life histories juvenile survival was the main driver behind the patterns in total demographic variance and extinction risk. This agrees with a general finding in a variety of matrix models studied by Carslake et al. (2009). Together with our results, this suggests that the importance of juvenile survival may hold for a large number of species in nature. Exceptions to this pattern are, in a broad sense, short-lived life histories with low fecundity, where fecundity is most important, and very long-lived life histories with low fecundity where adult survival dominates stochastic dynamics. These suggestions are, however, dependent on the degree to which the population models represent biological reality (Carslake et al. 2009).

Overall, our results agree well with empirical patterns between demographic variance and life history traits in birds found by Sæther et al. (2004). They also found increased demographic variance with increasing fecundity and decreasing age of maturation, and peaks of demographic variance at intermediate levels of adult survival. In contrast to the empirical findings of Sæther et al. (2004) we found that demographic variance (and hence extinction risk) increased almost monotonically with adult survival for life histories with high age of maturation. However, late maturing life histories with low to moderate adult survival rates were absent in Sæther et al.'s (2004) data (because such bird species hardly exist) and therefore could not show up in their relationships. The strong similarities between the empirical findings in birds and the theoretical results of our generic model lend support to our modeling framework as a method for comparing extinction risks among life histories. Further tests of predictions from our model using other empirical data on demographic variance are needed to corroborate the generality of the framework.

Our study is limited to the effects of demographic stochasticity on extinction risk. In reality, however, extinction dynamics of populations are also determined by environmental variation, density-dependence, population size, population structure, population growth rate and interactions among these factors (Cohen 1979, Tuljapurkar and Orzack 1980, Tuljapurkar 1982a, Tuljapurkar 1982b, Tuljapurkar 1990, Lande 1993, Benton and Grant 1996, Lande 1998, Lande et al. 2003, Benton et al. 2006). Thus, the contribution of our study to the general knowledge of extinction dynamics is an enhanced understanding of how life histories of species of small populations affect demographic variance of population growth and thereby extinction risks. Our study concerned density-independent situations but indicates that the life-history-related patterns in demographic variance remains also under density dependence. For example, at low population density and negative density dependence positive population growth will reduce extinction risk greatly, but the relative patterns of demographic variance in relation to life history traits will remain. The importance of life history for extinction dynamics has also been shown to be significant in models analyzing temporal autocorrelation of environmental variation and density dependence (Halley 1996, Ruokolainen et al. 2009). It is obvious that life-history strategies significantly interact with stochastic processes and density dependence in shaping extinction dynamics of small populations. Population structure should therefore always be considered both in generic studies of extinction dynamics and in management of threatened species.

## ACKNOWLEDGEMENTS

We wish to thank D. Arlt, T. Benton, C. Björkman, M. Low, T. Pärt and four anonymous reviewers for valuable comments on earlier versions of this manuscript, and M. Fujiwara for being helpful in clarifying some of the methods used.

## REFERENCES

Beissinger, S. R., and D. R. McCullough. 2002. Population viability analysis. University of Chicago Press, Chicago, U.S.A.

Beissinger, S. R., and M. I. Westphal. 1998. On the use of demographic models of population viability in endangered species management. Journal of Wildlife Management 62:821-841.

Benton, T. G., and A. Grant. 1996. How to keep fit in the real world: elasticity analyses and selection pressures on life histories in a variable environment. American Naturalist 147:115-139.

Benton, T. G., S. J. Plaistow, and T. N. Coulson. 2006. Complex population dynamics and complex causation: devils, details and demography. Proceedings of the Royal Society B-Biological Sciences 273:1173-1181.

Carslake, D., S. Townley, and D. J. Hodgson. 2009. Patterns and rules for sensitivity and elasticity in population projection matrices. Ecology 90:3258-3267.

Caswell, H. 2001. Matrix population models - construction, analysis, and interpretation. Second edition edition. Sinauer Associates, Inc. Publishers, Sunderland, MA, USA.

Charlesworth, B. 1994. Evolution in age-structured populations. Cambridge University Press, Cambridge.

Cohen, J. E. 1979. Comparative statics and stochastic dynamics of age-structured populations. Theoretical Population Biology 16:159-171.

Drake, J. M. 2005. Density-dependent demographic variation determines extinction rate of experimental populations. PLoS Biology 3:1300-1304.

Engen, S., R. Lande, B. E. Sæther, and H. Weimerskirch. 2005. Extinction in relation to demographic and environmental stochasticity in age-structured models. Mathematical Biosciences 195:210-227.

Fujiwara, M. 2007. Extinction-effective population index: incorporating life-history variations in population viability analysis. Ecology 88:2345-2353.

Fujiwara, M., and H. Caswell. 2001. Demography of the endangered North Atlantic right whale. Nature 414:537-541.

Gabriel, W., and R. Burger. 1992. Survival of small populations under demographic stochasticity. Theoretical Population Biology 41:44-71.

Gabriel, W., and R. Ferrière. 2004. From individual interactions to population viability. Pages 19-40 *in* R. Ferrière, U. Dieckmann, and D. Couvet, eds. Evolutionary Conservation Biology. Cambridge University Press, Cambridge, UK.

Gilpin, M. 1992. Demographic stochasticity: A Markovian approach. Journal of Theoretical Biology 154:1-8.

Halley, J. M. 1996. Ecology, evolution and 1/f-noise. Trends in Ecology & Evolution 11:33-37.

Harris, T. H. 1963. The theory of branching processes. Springer-Verlag, Berlin.

Jonsson, A., and B. Ebenman. 2001. Are certain life histories particularly prone to local extinction? Journal of Theoretical Biology 209:455-463.

Keiding, N. 1975. Extinction and exponential growth in random environments. Theoretical Population Biology 8:49-63.

Kendall, B. E., and G. A. Fox. 2002. Variation among individuals and reduced demographic stochasticity. Conservation Biology 16:109-116.

Kendall, B. E., and G. A. Fox. 2003. Unstructured individual variation and demographic stochasticity. Conservation Biology 17:1170-1172.

Kendall, B. E., and M. E. Wittmann. 2010. A stochastic model for annual reproductive success. American Naturalist 175:461-468.

Kokko, H., and T. Ebenhard. 1996. Measuring the strength of demographic stochasticity. Journal of Theoretical Biology 183:169-178.

Kraus, C., D. L. Thomson, J. Künkele, and F. Trillmich. 2005. Living slow and dying young? Life-history strategy and age-specific survival rates in a precocial small mammal. Journal of Animal Ecology 74:171-180.

Lande, R. 1993. Risks of population extinction from demographic and environmental stochasticity and random catastrophes. American Naturalist 142:911-927.

Lande, R. 1998. Demographic stochasticity and Allee effect on a scale with isotropic noise. Oikos 83:353-358.

Lande, R. 2002. Incorporating stochasticity in population viability analysis. Pages 18-40 *in* S. R. Beissinger and D. R. McCullough, eds. Population viability analysis. The University of Chicago Press, Chicago.

Lande, R., S. Engen, and B. E. Sæther. 2003. Stochastic population dynamics in ecology and conservation. Oxford University Press, Oxford.

Lee, A. M., B. E. Sæther, and S. Engen. 2011. Demographic stochasticity, Allee effects, and extinction: the influence of mating system and sex ratio. American Naturalist 177:301-313.

Legendre, S., J. Clobert, A. P. Møller, and G. Sorci. 1999. Demographic stochasticity and social mating system in the process of extinction of small populations: the case of passerines introduced to New Zealand. American Naturalist 153:449-463.

Leigh, E. G. 1981. The average lifetime of a population in a varying environment. Journal of Theoretical Biology 90:213.

Melbourne, B. A., and A. Hastings. 2008. Extinction risk depends strongly on factors contributing to stochasticity. Nature 454:100-103.

Mode, C. J., and G. T. Pickens. 1986. Demographic stochasticity and uncertainty in population projections - a study by computer simulation. Mathematical Biosciences 79:55-72.

Morris, W. F., and D. F. Doak. 2002. Quantitative conservation biology: theory and practice of population viability analysis. Sinauer Associates Inc., Sunderland, MA, U.S.A.

Morris, W. F., and D. F. Doak. 2004. Buffering of life histories against environmental stochasticity: accounting for a spurious correlation between the variabilities of vital rates and their contributions to fitness. American Naturalist 163:579-590.

Oli, M. K. 2003a. Partial life-cycle analysis: a simplified model for post-breeding census data. Ecological Modelling:101-108.

Oli, M. K. 2003b. Partial life-cycle models: how good are they? Ecological Modelling:313-325.

Oli, M. K., and B. Zinner. 2001. Partial life-cycle analysis: A model for birth-pulse populations. Ecology 82:1180-1190.

Pimm, S. 1991. The balance of nature. The University of Chicago Press, Chicago.

Richter-Dyn, N., and N. S. Goel. 1972. On the extinction of a colonizing species. Theoretical Population Biology 3:406-433.

Ruokolainen, L., A. Lindén, V. Kaitala, and M. S. Fowler. 2009. Ecological and evolutionary dynamics under coloured environmental variation. Trends in Ecology & Evolution 24:555.

Sæther, B. E., and O. Bakke. 2000. Avian life history variation and contribution of demographic traits to the population growth rate. Ecology 81:642-653.

Sæther, B. E., S. Engen, A. P. Møller, M. E. Visser, E. Matthysen, W. Fiedler, M. M. Lambrechts, P. H. Becker, J. E. Brommer, J. Dickinson, C. Du Feu, F. R. Gehlbach, J. Merila, W. Rendell, R. J. Robertson, D. Thomson, and J. Torok. 2005. Time to extinction of bird populations. Ecology 86:693-700.

Sæther, B. E., S. Engen, A. P. Møller, H. Weimerskirch, M. E. Visser, W. Fiedler, E. Matthysen, M. M. Lambrechts, A. Badyaev, P. H. Becker, J. E. Brommer, D. Bukacinski, M. Bukacinska, H. Christensen, J. Dickinson, C. du Feu, F. R. Gehlbach, D. Heg, H. Hotker, J. Merilä, J. T. Nielsen, W. Rendell, R. J. Robertson, D. L. Thomson, J. Torok, and P. Van Hecke. 2004. Life-history variation predicts the effects of demographic stochasticity on avian population dynamics. American Naturalist 164:793-802.

Shaffer, M. 1987. Minimum viable populations: coping with uncertainty. Pages 69-86 *in* M. E. Soulé, ed. Viable populations for conservation. Cambridge University Press, Cambridge.

Shaffer, M. L. 1981. Minimum population sizes for species conservation. Bioscience 31:131-134.

Silvertown, J., M. Franco, and E. Menges. 1996. Interpretation of elasticity matrices as an aid to the management of plant populations for conservation. Conservation Biology 10:591-597.

Soulé, M. E., and B. A. Wilcox. 1980. Conservation Biology: An evolutionary-ecological perspective. Sinauer Associates., Sunderland, MA.

Tuljapurkar, S. 1990. Population dynamics in variable environments. Springer-Verlag, New York.

Tuljapurkar, S. D. 1982a. Population dynamics in variable environments. II. Correlated environments, sensitivity analysis and dynamics. Theoretical Population Biology 21:114.

Tuljapurkar, S. D. 1982b. Population dynamics in variable environments. III. Evolutionary dynamics of r-selection. Theoretical Population Biology 21:141-165.

Tuljapurkar, S. D., and S. H. Orzack. 1980. Population dynamics in variable environments I. Long-run growth rates and extinction. Theoretical Population Biology 18:314-342.

Vindenes, Y., S. Engen, and B. E. Sæther. 2008. Individual heterogeneity in vital parameters and demographic stochasticity. American Naturalist 171:455-467.

# Appendix

**Figure A1:** The relationship between extinction risk and demographic variance, including all simulated life histories listed in appendix B, available online and shown in figs. 1–3. Extinction risk is estimated with the branching process, and demographic variance with equation (5).

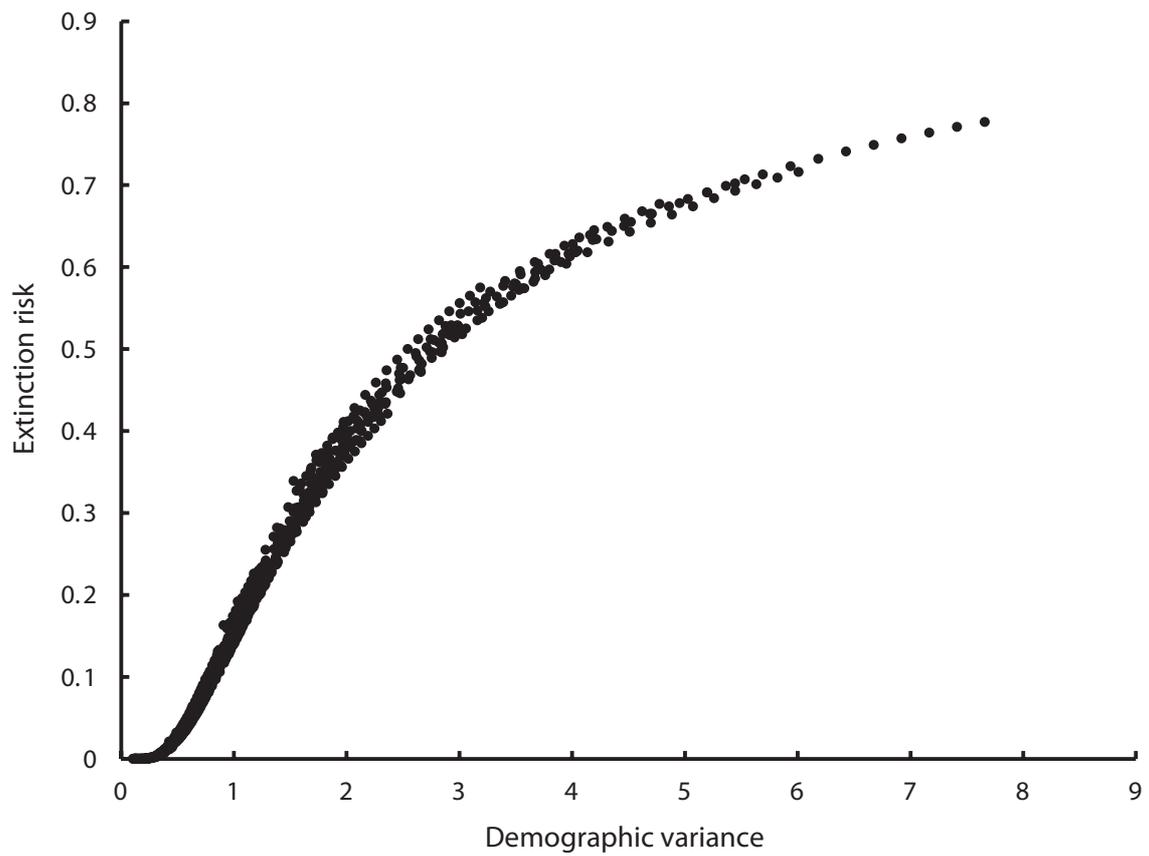

**Figure A2:** Relationships between life-history traits and sensitivity. For calculation of sensitivities, see "Methods". For juvenile survival only sensitivities of the first juvenile stage is shown. The figures show the same selection of life histories as in figure 5. $F_j$ sometimes overlaps with $F_a$, and is in these cases not visible in the figures. A legend common to all graphs is shown in the first subfigure.

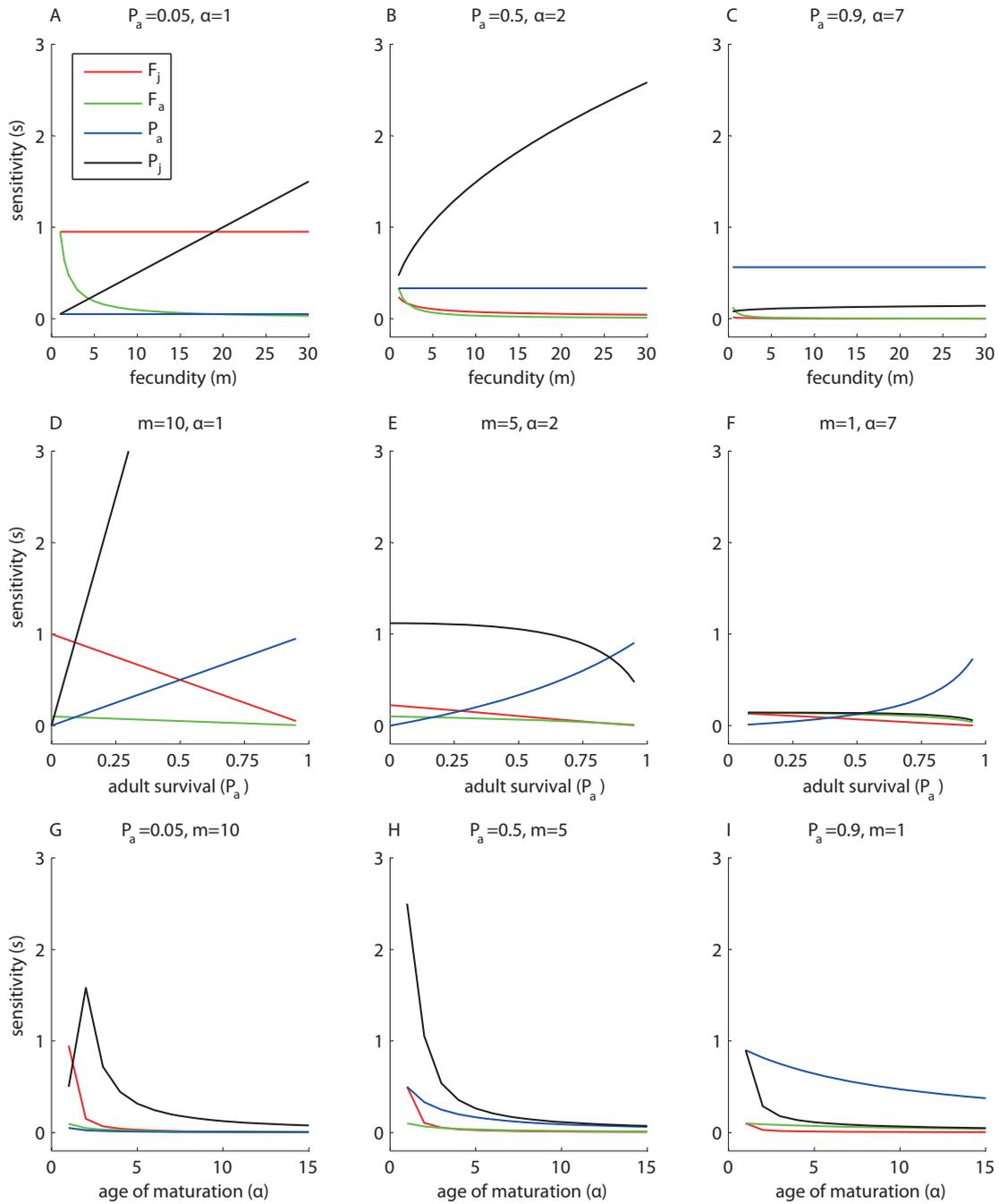

**Figure A3:** Relationships between life history traits and variance of stage specific survival and fertility. Variances are calculated for matrix entries, that is, $F_j$, $F_a$, $P_j$, and $P_a$, according to assumed probability distributions (Poisson distribution for $F_x$ and binomial distribution for $P_x$; see "Methods"). The figures show the same selection of life histories as in figure 5. $F_j$ sometimes overlaps with $F_a$, and is in these cases not visible in the figures. A legend common to all graphs is shown in the first subfigure.

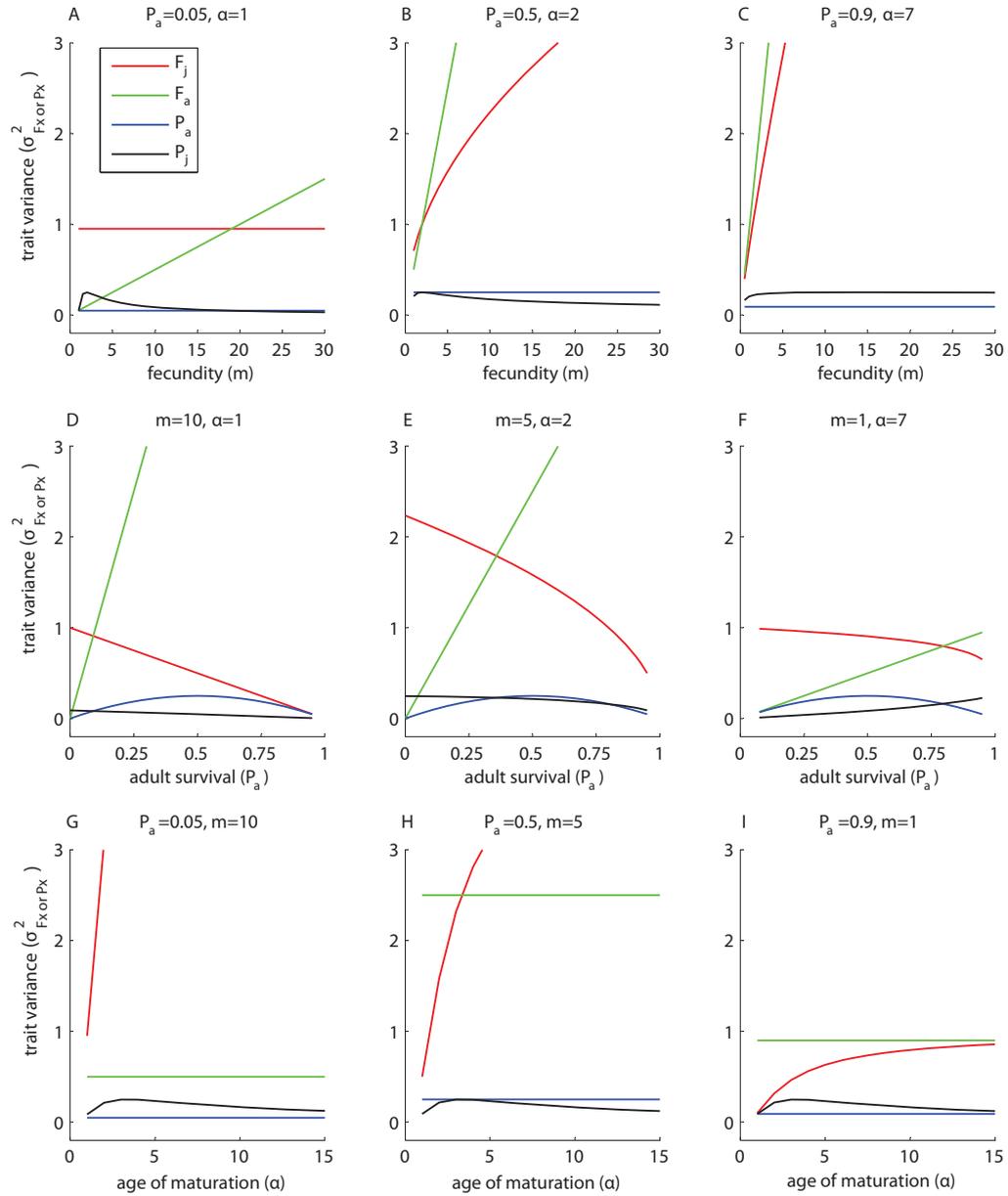

Appendix B, available online.

See separate file.